# Nanophotonic control of Forster Resonance Energy Transfer

Niels Zijlstra*, Allard P. Mosk[#], Willem L. Vos,[#] Vinod Subramaniam*, and Christian Blum*

*Nanobiophysics, Faculty of Science and Technology and MESA+ Institute for Nanotechnology, University of Twente, P.O. Box 217, 7500 AE Enschede, The Netherlands.*

*# Complex Photonic Systems (COPS), Faculty of Science and Technology and MESA+ Institute for Nanotechnology, University of Twente, P.O. Box 217, 7500 AE Enschede, The Netherlands.*

Here we report on the experimental details of a study on the influence of the photonic environment on the emission of a FRET system. We modified the local density of optical states (LDOS) by placing the FRET system at precisely defined distances to a metallic mirror. We measured the energy donor lifetime in the presence of the acceptor, $\tau_{DA}$, of the FRET system and the lifetime $\tau_D$ of an identical sample lacking an acceptor fluorophore for different LDOS. Since energy transfer is effectively an additional decay channel for the donor, the decay rates $\gamma_D = \frac{1}{\tau_D}$ and $\gamma_{DA} = \frac{1}{\tau_{DA}} = \gamma_D + \gamma_{FRET}$ give access to the energy transfer rate,

$$\gamma_{FRET} = \gamma_{DA} - \gamma_D \qquad (1)$$

and the energy transfer efficiency [1]

$$E_{FRET} = 1 - \frac{\gamma_D}{\gamma_{DA}} \qquad (2)$$

for the specific value of LDOS sampled.

**Fluorophore-DNA FRET pair and donor only reference**

We used the fluorescent dyes Atto488 and Atto565 as donor and acceptor fluorophores, respectively. The dyes were separated by a single DNA strand of 15 base pairs: Atto488 - CGA CTC CGA GTC AGC – Atto565. The unlabeled complementary strand was used to create a double-stranded DNA construct, assuring that the fluorophores were rigidly separated on the DNA scaffold.

The HPLC and PAGE purified single stranded DNA was obtained from IBA Biotagnology (Germany). The double stranded DNA was formed by mixing the labeled DNA, the unlabeled complementary strand, and buffer (5mM Tris, 100mM NaCl). The mixture was heated to 80°C for 10 minutes and then annealed by slowly cooling to room temperature. The donor-only reference sample, that is in the absence of the FRET acceptor, consisted of atto488 covalently attached to an identical double strand of DNA (atto488-DNA) and was annealed in an identical way as the double labelled FRET strand.

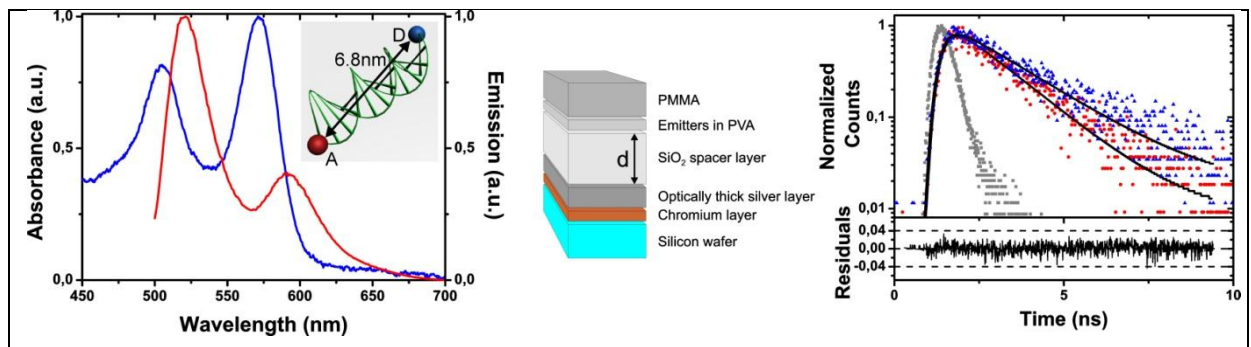

Figure 1( a) Solution absorbance (blue) and emission spectra (red) of the analyzed FRET system consisting of a short double-stranded DNA labeled with the donor atto488 (D) and acceptor atto565 (A) fluorophores respectively. (b) Schematic of the sample design used to modulate the LDOS. We deposited a $SiO_2$ spacer layer of well-defined thickness in the range 60 nm to 270 nm on a silver layer that acts as a mirror. On this spacer layer a thin (<20 nm) PVA layer containing the emitters was spincoated and covered with a thick layer of PMMA of matching refractive index to $SiO_2$. (c) Typical decay curves measured for the FRET sample atto488-DNA-atto565 in absence (reference sample, blue triangles, τ = 1.80 ns) and presence of a mirror modulating the LDOS at energy donor frequencies (red circles, d = 60 nm, τ = 1.39 ns). The curves are normalized to unity, the light grey squares correspond to the instrument response function (IRF) of the system and the black lines correspond to single exponential fits. The residuals for the reference sample are plotted at the bottom.

**Controlling the LDOS**

To control the LDOS we placed the emitters at precisely defined distances from a metallic mirror. The presence of the mirror results in an oscillation of the LDOS as a function of distance to the mirror that further results in an oscillation of the emission rate of the fluorophore measured.. In this way control over the distance from the emitters to the mirror corresponds to control of the LDOS. To model our data we used a classical model taking full account of material properties of the mirror and dielectric environment developed by Chance et al [2].

**Sample preparation**

We briefly describe the fabrication of planar single mirror samples (for details see [3]). First, the chromium, silver, and silicon oxide layers were fabricated using multilayer e-gun deposition on a silicon wafer, carried out in a Balzers BAK 600 e-gun evaporation machine. The thickness of the silicon oxide layer was varied to control the distance between the mirror layer and the emitters. The thickness of the layers was determined using scanning electron microscopy (SEM) within an estimated error of 5%. The thickness of the layers was found to be homogeneous over the analyzed cross sections. The refractive index of $SiO_2$ was determined to n=1.46 ± 0.05 by white light ellipsometry.

We then spincoated a thin layer of polyvinyl alcohol (PVA, n=1.50 ± 0.01) over the $SiO_2$ spacer layer. The PVA was 2% by weight dissolved in water and contained the DNA FRET samples at low concentrations (~1 μM). The estimated distance between individual FRET pairs is 25 nm so that interaction between FRET systems could be neglected. We spincoated the PVA layer at 6000 rpm for 10 seconds, resulting in a homogeneous layer thickness of 17 ± 3 nm. Finally, a thick (>1 μm) layer of polymethyl methacrylate (PMMA, n=1.49 ± 0.01) was spincoated at 2800 rpm for 10 seconds on top of the PVA layer to avoid reflections from the PVA/air interface and to create an optically isotropic environment.

Different emitter-mirror distances, and hence different LDOS, were achieved by varying the thickness of the deposited $SiO_2$ spacer layer between 60 nm to 270 nm (schematic see figure 1b). Additionally we fabricated a reference sample without mirror (that is, with the mirror at infinite distance).

**Instrumentation**

Time-correlated single-photon counting (TCSPC) measurements were performed using a custom-built inverted confocal fluorescence lifetime imaging microscope. As excitation source, we used a pulsed diode laser operating at 485nm at a repetition rate of 20 MHz (LDH-D-C-485, Picoquant). An epi-illumination configuration was used: the sample was illuminated and the emission was collected through the same microscope objective (UPLSAPO 60XW, 60X, 1.2NA, Olympus). Any remaining excitation light in the detection path was blocked with a long pass filter (RazorEdge, 488 nm, Semrock) and an additional single-notch filter (Stopline, 488/14 nm, Semrock). To perform time-resolved fluorescence measurements, the emission was focused onto the active area of a single photon avalanche diode (SPCM-APQR-16, PerkinElmer), connected to the TCSPC module (PicoHarp300, Picoquant). To limit the detection wavelength to the emission peak of the FRET donor fluorophore, we used a band pass filter centered at 525 nm with a bandwidth of 15 nm (BrightLine,

525/15 nm, Semrock). The integration time per decay curve was chosen such that the total counts per decay curve exceeded 20000 counts to assure accurate determination of the characteristic lifetime [4].

**Measurement procedure**

We scanned three areas of 50 by 50 μm and collected an average of 400 decay curves per area, see figure 2. Each decay curve was fitted with a single-exponential decay. The quality of the fit was determined by the reduced chi-square parameter. From the lifetimes determined, we obtained one histogram per area. The distribution of lifetimes is fitted with a Gaussian distribution. A narrow distribution indicates a good sample quality. Small variations in the refractive index and in the thickness of the spacer layer will result in small deviations in the measured decay rate. If these variations are too large due to poor sample quality, the distribution will be broad compared to the most frequent lifetime.

The decay rate was calculated as the inverse of the peak value of the lifetime distribution, whereas the error for the decay rate was derived from the width of the distribution, typically $\Delta\tau=0.09$ ns. From the calculated decay rate and width of the distribution measured independently on the three areas, we determined the average donor fluorophore decay rate and error for each specific value of the LDOS, for both samples containing the complete FRET system (i.e. donor and acceptor) and a control sample containing a donor fluorophore alone in the absence of the acceptor.

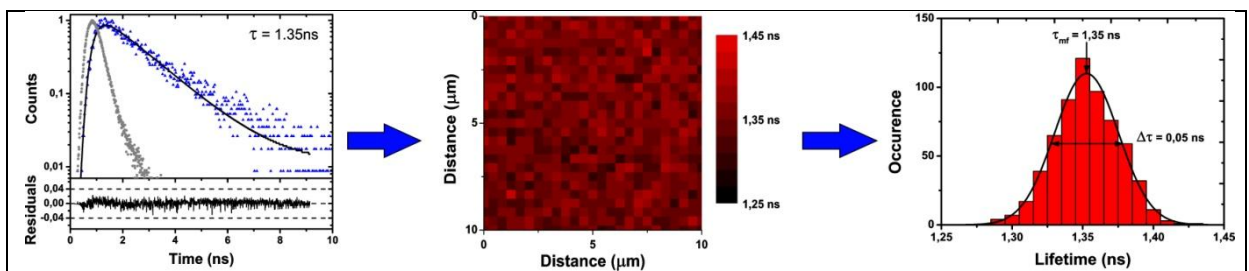

Figure 2: **(a)** Typical decay curve measured for the FRET donor at 60nm from a silver mirror (blue triangles), measured at λ = 525 ± 15 nm. The solid black line is the single-exponential fit to the data. The grey circles correspond to the instrument response function (IRF) of the system. The residuals are plotted at the bottom and show the good quality of the fit in the whole decay range. (b) Typical lifetime image, showing a uniform distribution of the lifetimes. Each pixel represents the lifetime extracted from a fit similar to the one shown in (a).(c) The lifetime histogram obtained from the lifetime image in (b). The histogram shows a narrow Gaussian distribution indicating a good sample quality.

**Samples analyzed**

We prepared 8 different mirror platforms with SiO$_2$ spacer layers between 60 nm to 270 nm plus one reference sample without mirror, which gave us in total 9 different specific LDOS values. For each spacer thickness, representing one LDOS value, one sample consisting of the donor only atto488-DNA and one FRET pair sample of atto488-DNA-atto565 were prepared. The fluorescence lifetime of the donor atto488 fluorophore in a complete FRET pair was determined for all LDOS values, as was the fluorescence lifetime of donor only control samples as detailed above.

We thus obtained a dataset that contained the precise change of the FRET donor decay rate $\gamma_{DA}$ (inverse of measured lifetime $\tau_{DA}$) and donor only decay rate $\gamma_D$ (inverse of measured lifetime $\tau_D$) in the absence of the FRET acceptor upon changing the LDOS. Using equation 1 and 2 we access the dependence of the FRET rate $\gamma_{FRET}$ and the FRET efficiency $E_{FRET}$ as a function of the LDOS at donor frequencies.

The detailed results of our study showing the relation between LDOS and $\gamma_{FRET}$ and LDOS and $E_{FRET}$ will be published shortly.